\documentclass[runningheads,a4paper]{llncs}

\usepackage{amssymb}
\usepackage{graphicx}

\usepackage{url}

\urldef{\mailsa}\path| rein.prank@ut.ee|

\begin{document}

\mainmatter  

\title{Software for Evaluating Relevance of Steps in Algebraic Transformations\thanks{  The final publication is available at http://link.springer.com.}}

\titlerunning{Relevance of steps in algebraic transformations}

%
\author{Rein Prank}

\authorrunning{R. Prank}

\institute{University of Tartu, Estonia
}

\toctitle{Lecture Notes in Computer Science}
\tocauthor{Authors' Instructions}
\maketitle

\begin{abstract}
Students of our department solve algebraic exercises in mathematical logic in a computerized environment. They construct transformations step by step and the program checks the syntax, equivalence of expressions and completion of the task. With our current project, we add a program component for checking relevance of the steps. 
\end{abstract}

\section{Introduction}

Computerized exercise environments for algebraic transformations try to pre-
serve equivalence of expressions but they usually do not evaluate whether solu-
tion steps are relevant (for the actual task type) or not. Some versions of Algebra Tutors of Carnegie Mellon University in the nineties required a prescribed solution path to be followed. For example, Equation Solving Tutor 
 \cite{Ritt} counted division before subtraction in $2x=11-3$ as an error. But the review article ``Cognitive Tutors: Lessons Learned'' \cite{Less} summarizes: ``Our earlier tutors required students to always stay on path. More recent tutors allow the student to go off path but still focus instruction on getting student back on path ...''.  There is one commonly known algebra environment,  Aplusix \cite{Nic}, where the program displays the ratios of what part of the syntactic goals \emph{factored}, \emph{expanded}, \emph{reduced}, \emph{sorted} is already reached and what part remains. However the ratios in itself are not of much help for a student. For example, if the student does not reduce the fraction $ba/bc$ but converts it to $ab/bc$ then the ratios simply indicate some improvement with regard to the goal \emph{sorted}. 

Students of our department have solved technical exercises in Mathematical Logic on computers since 1991. One of our programs is an environment for algebraic transformations \cite{PraVi,PraVa}. For many years it seemed that checking of syntax, order of logical operations and equivalence of expressions is sufficient for training and assessment. Some years ago the introductory part of propositional logic containing also tasks on expressing of given formulas using $\{\&,\lnot\}$, $\{\lor,\lnot\}$ or $\{\supset,\lnot\}$ only and on disjunctive normal form (DNF) was moved into the first-term course 'Elements of Discrete Mathematics'. We saw that, besides students who solved our exercises very quickly, there were others who were in real trouble. Most problematic were DNF exercises where many solutions had a length of 50--70 steps or more. The instructors were not able to analyze long solutions (note that the main program does not record the marking and conversion rule but only displays rows with formulas). We decided to write an additional program that checks the relevance of solution steps and annotates the solutions.

Our main program, analysis tool and some other necessary files are available at \url{http://vvv.cs.ut.ee/~prank/rel-tool.zip}. The paper describes the basic environment (Section 2) and our supplementary tool for normal form exercises (Section 3). Section 4 provides some discussion of further opportunities. 

\section{Correctness Checking in the Main Program}

Working in our formula transformation environment, the student creates the solution step by step. Each conversion step consists of two substeps. At the first substep the student marks a subformula to be changed. For the second substep the program has two different modes. In the INPUT mode the program opens an input box and the student enters a subformula that replaces the marked part. In the RULE mode the student selects a rule from the menu and the program applies it. Figure 1 demonstrates a DNF exercise in the RULE mode. 

\vspace{-0.2cm}
\begin{figure}
\centering
\includegraphics [scale=.64] {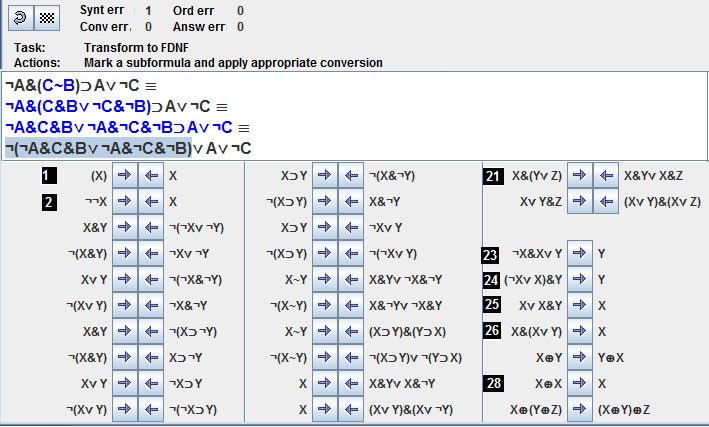}
\caption{Solution window of the main program. The student has performed three steps and marked a subformula for moving the negation inside.}
\end{figure}
\vspace{-0.5cm}

At the first substep the program checks whether the marked part is a proper subformula. At the second substep in the INPUT mode the program checks syntactical correctness of the entered subformula and equivalence. In the RULE mode the program checks whether the selected rule is applicable to the marked part. In case of an error the program requires correction. However, our main program does not evaluate the relevance of conversions. 

In our course the exercises on expression of formulas using given connectives are solved in the INPUT mode and exercises on DNF in the RULE mode. 

\section{A Tool for Solution Analysis}

Our lectures contain the following six-stage version of the algorithm for conversion of formulas to full disjunctive normal form:
\vspace{-0.2cm}
\begin{enumerate}
\item Eliminate implications and biconditionals from the formula. 
\item Move negations inside. 
\item Use distributive law to expand the conjunctions of disjunctions.
\item Exclude contradictory conjunctions and redundant copies of literals. 
\item Add missing variables to conjunctions.
\item Order the variables alphabetically, exclude double conjunctions.
\end{enumerate}

We now describe how the analysis tool treats relevance of solution steps. The program accepts the choice of the rule if it corresponds to the algorithm stage or is one of the simplification rules (rules in positions 1--2, 23--26 and 28 in Figure 1). For some conversions the tool checks additionally that the rule is applied reasonably. Elimination of biconditional should not duplicate implications and biconditionals. Negations should be moved inside starting from the outermost negation. All the literals of a conjunction should be ordered alphabetically in one step. (There are some more checks of similar type).
 
The analysis tool displays on the screen and records in a text file for each step an annotation that contains the following information:
\vspace{-0.2cm}
\begin{enumerate}
\item Number of the stage in the FDNF algorithm [+ a clue about the conversion]. 
\item Number and meaning of the applied rule + OK if the step was acceptable. 
\item Error message if the step was not acceptable.
\item Initial and resulting formula with the changed/resulting part highlighted. 
\end{enumerate}

For example, the five lines below will be recorded as the annotation of solution step 2 in Figure 1. The symbol '$\gg$' denotes implication. Rectangles and triangles point to the changed part of the formula and to the error message.
\vspace{-0,5cm}
\begin{figure}
\includegraphics [scale=.5] {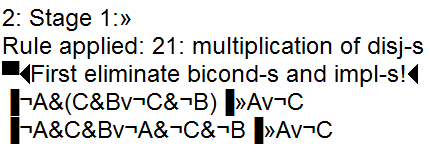}
\end{figure}

\vspace{-0.5 cm}
The tool also compiles statistics of error messages in the whole solution file of the student and statistics of the group of students. This statistics is recorded in the form of tables where the rows correspond to separate solution attempts of the tasks and the columns are for particular error types and for some other characteristics (number of steps, number of steps taken back, total number of errors, stage reached in the solution algorithm). This output can be copied into a spreadsheet environment for further statistical treatment.  

The analysis tool gives an error message when the formula contains independent parts that are in different stages of the algorithm and the applied conversion does not correspond to the stage of the whole formula. However, in such cases it is quite easy to understand whether the step makes the solution longer or not. 

Does the tool find all reasons for long solutions? Our initial count of the possibilities for rule misapplication gave us 15 error types for full DNF tasks. A comparison of solutions and received annotations disclosed several additional unwise approaches to performing the \emph{right} conversions: incomplete reordering of variables, addition of one variable instead of two etc. After including them we ended up with 19 error types. The most frequent errors are presented in Table 1. 

From the scanned solution files we learned about a further, `more delicate' solution economy problem. The algorithm prompts the user to apply the distributivity law at stage 3 and to eliminate redundant members at stage 4. Such ordering enables a very straightforward proof of the feasibility of the algorithm. However, it is often useful to perform some conversions of stage 4 before stage 3. Our analyzer does not require nor prohibit this. Conversions of stage 4 use only simplification rules and they do not evoke error messages.  

\begin{table}
\caption{Results and numbers of diagnosed errors in final tests in 2011 and 2012 }
\begin{center}
\setlength\tabcolsep{5pt}
\begin{tabular}{|l|r|r|}
\hline
$\mathbf{Quantity/error}$ & $\mathbf{Test\ 2011}$ & $\mathbf{Test\ 2012}$\\
\hline
 Number of solutions (completed/total) & 131/162 & 150/169\\
\hline
Steps &	5766/7270 & 4096/4764\\
\hline
Steps taken back & 500/933 & 112/186\\
\hline
Relevance errors diagnosed & 1097/1481 & 321/502\\
\hline
4. Negation moved into brackets at stage 1 & 39/56 & 32/33\\
\hline
5. Negation moved out of brackets & 55/157 & 12/36\\
\hline
6. Inner negation processed first & 142/219 & 68/94\\
\hline
7. Distributive law applied too early & 70/84 & 35/44\\
\hline
9. Members reordered too early & 274/327 & 21/52\\
\hline
10. Members of FALSE conjunction reordered & 67/81 & 13/39\\
\hline
11. Reordering together with redundant members & 59/68 & 12/13\\
\hline
12.Members of disjunction reordered (as for CNF) & 192/197 & 32/33\\
\hline
13. Variables added too early & 102/147 & 45/74\\
\hline
16. Only a part of conjunction reordered & 45/45 & 15/16\\
\hline
Average number of steps in completed solutions & 44.0 & 27.3\\
\hline
\end{tabular}
\end{center}
\end{table}
 
\vspace{-0.5cm} 
Table 1 presents data about solutions of a full DNF task in the final tests of 2011 and 2012. Randomly generated initial formulas contained four different binary connectives and 2--3 negations (like Fig. 1). The results of 2011 looked rather disappointing. With 185 students taking the test, 162 of them submitted the solution file of formula transformation tasks, and the full DNF task was completed in 131 files. The average number of steps in completed solutions was 44 when the optimal number was 15--25. Very often several steps had been taken back (using Undo).  

In the autumn term of 2012 we made the analyzer available to the students, although it does not have a developed user interface. We added a small task file with only two full DNF tasks and required that they submit a solution file where each of the two solutions can only contain one diagnosed relevance error. The students could also use the annotation tool when preparing for the final test. The results of 2012 in Table 1 demonstrate that the annotation tool is useful for the students as well. 

\section{Extending the Approach to Other Situations}

It seems that our current program is able to produce satisfactory explicit diagnosis of the relevance of steps in solutions of DNF and CNF tasks in RULE mode. There is an obvious extension to the algorithmically less interesting tasks on expression of formulas using negation and one binary connective.

Is it possible to apply relevance checking to the conversions in INPUT mode? Our students solve some exercises in INPUT mode. The relevance tool is designed to determine what rule is used for the step and so we had the opportunity to scan the input-based solutions. We discovered that virtually all steps were performed using the same rules 1--29, sometimes removing double negations from the result of the step. Nevertheless it is clear that for understanding free conversions we should replace the indirect identification of a single rule by direct modelling of one or more sequentially applied rules. It probably also means replacing our string representations with structured representations of mathematical objects and using the tools that work in these representations.

There exists a very powerful rule-based conversion environment, Mathpert,  for algebra and calculus exercises \cite{Bees} (later versions are called MathXpert).  It could be a quite interesting task to complement MathXpert with relevance checking. 

\subsubsection*{Acknowledgments.} Current research is supported by Targeted Financing grant  SF0180008s12 of the Estonian Ministry of Education. 


\end{document}